\documentclass{webofc}
\usepackage[varg]{txfonts}
\usepackage{bm}
\begin{document}
\title{Holographic description of total hadronic cross sections at high energies}
\author{\firstname{Akira} \lastname{Watanabe}\inst{1,2}\fnsep\thanks{\email{akira@ihep.ac.cn}} \and
        \firstname{Mei} \lastname{Huang}\inst{1,2}\fnsep\thanks{\email{huangm@ihep.ac.cn}}
}

\institute{
Institute of High Energy Physics and Theoretical Physics Center for Science Facilities, Chinese Academy of Sciences, Beijing 100049, People's Republic of China
\and
University of Chinese Academy of Sciences, Beijing 100049, People's Republic of China
}

\abstract{
We investigate the hadron-hadron total cross sections at high energies in the framework of holographic QCD.
In our model setup, the involved strong interaction is described by the Brower-Polchinski-Strassler-Tan Pomeron exchange kernel, and the two form factors are obtained from the bottom-up AdS/QCD models.
We show that the resulting nucleon-nucleon total cross section is consistent with the experimental data recently measured by the TOTEM collaboration at the LHC.
As examples to see the general versatility of the model, we also present our analysis on the pion-nucleon and pion-pion cases, which can be predicted because all the adjustable parameters are fixed by the nucleon-nucleon data.
}
\maketitle
%
\section{Introduction}
\label{sec-1}
Understanding the partonic structures of the hadrons is one of the most important subjects in hadron physics.
High energy scattering phenomena, which can be measured experimentally, have played crucial roles in revealing them for many years.
The strong interaction in the scattering processes gives a dominant contribution to the measured cross sections, and the fundamental theory to describe it is knows as quantum chromodynamics (QCD).
The expression for the cross section can be factorized into two components, the ``hard'' (perturbative) and the ``soft'' (nonperturbative) parts.
The perturbative technique in QCD is applicable to perform analysis on the former one, but cannot used for the latter one.
The soft part is described by the parton distribution functions (PDFs) which are expressed by the Bjorken variable $x$ and the probe energy scale $Q^2$ and parameterized with experimental data from various scattering processes.
Since the PDFs are not calculable quantities, they are in principle treated as inputs in any analysis of high energy processes.

However, particularly in the small $x$ region, our knowledge is still quite poor and the parameterized PDFs have huge uncertainties, which sometimes make the predictions unreliable, because it is not easy to correctly evaluate such uncertainties in the final results.
In this kinematic region, the strong interaction among so many gluons with tiny momenta gives a dominant contribution to the cross section.
Since precisely considering such an extremely complicated dynamics based on QCD is impossible in practice, it is much more suitable to perform the analysis by effective approaches.
The Pomeron exchange is well-known as an effective description of the multi-gluon exchange which may be occurred in various scattering processes in the small $x$ region~\cite{ForshawRoss,PomeronPhysicsandQCD}.
This is practically the only way to treat the complicated dynamics, and has been successfully applied to the analysis of various processes such as the hadron-hadron scattering~\cite{Donnachie:1992ny} and the deep inelastic scattering (DIS).

The effective description by the Pomeron exchange is realized also in holographic QCD which is an effective approach constructed based on the anti-de Sitter/conformal field theory (AdS/CFT) correspondence (or gauge/string correspondence in general)~\cite{Maldacena:1997re:Gubser:1998bc:Witten:1998qj}.
This effective method in QCD has gathered a lot of theoretical interests~\cite{Kruczenski:2003be,Son:2003et,Kruczenski:2003uq,Sakai:2004cn:Sakai:2005yt,Erlich:2005qh,DaRold:2005zs}, and focusing mainly on DIS, many related studies have been done so far~\cite{Polchinski:2001tt:Polchinski:2002jw,BoschiFilho:2005yh,Brower:2006ea,Hatta:2007he,Brower:2007qh,BallonBayona:2007rs,Brower:2007xg,Cornalba:2008sp,Pire:2008zf,Cornalba:2010vk,Brower:2010wf,Watanabe:2012uc:Watanabe:2013spa,Watanabe:2015mia}.
In holographic QCD, the Pomeron in the usual Minkowski space is identified as the Reggeized graviton in the higher dimensional curved space, and one of the most successful descriptions of the holographic Pomeron was proposed by Brower-Polchinski-Strassler-Tan (BPST)~\cite{Brower:2006ea}.
The most remarkable result with the BPST kernel may be that the nontrivial scale dependence of the Pomeron intercept, which were measured at HERA, can be well reproduced by applying this kernel to calculate the structure function at small $x$~\cite{Brower:2010wf,Watanabe:2012uc:Watanabe:2013spa}.

In this report, based on Ref.~\cite{Watanabe:2018owy} we present our recent analysis on total cross sections of the high energy hadron-hadron scattering in the framework of holographic QCD.
So far, the DIS has intensively been studied, in which the density distribution of the probe photon is localized around the ultraviolet (UV) boundary in the AdS space, while the peak position of the target hadron density distribution is located in the infrared (IR) region.
Hence, the DIS is a UV-IR scattering, and the extracted Pomeron intercept depends on the four-momentum squared of the probe photon $Q^2$.
On the other hand, the hadron-hadron scattering is interpreted as an IR-IR scattering.
Since there is no such a variable scale in this scattering process, the Pomeron intercept becomes constant and its value may be close to that of the well-known soft Pomeron intercept 1.0808~\cite{Donnachie:1992ny}.
As to the proton-proton collision, the new experimental data at $\sqrt{s} = 13$~TeV, where $s$ is the Mandelstam variable, were taken recently by the TOTEM collaboration at the LHC~\cite{Antchev:2017dia}.
Since the only scale that the involved hadrons have in this process is the proton mass $m_p \sim 1$~GeV, the Bjorken $x$ value at $\sqrt{s} \sim 10$~TeV can be estimated as $x \sim 10^{-8}$ which is much smaller than that was realized at HERA.
Therefore, the new data provide us with valuable opportunities to investigate the very small $x$ region which cannot be studied enough before.

In this work, we firstly focus on the nucleon-nucleon case.
Employing the BPST Pomeron exchange kernel to describe the involved strong interaction between the two nucleons and the bottom-up AdS/QCD model of the nucleon~\cite{Hong:2006ta} to calculate its density distribution, which also is identified as the gravitational form factor~\cite{Abidin:2009hr}, in the five-dimensional AdS space, and then we obtain the total cross section at high energies.
We concentrate on the kinematic range $10^2 < \sqrt{s} < 10^5$~GeV, and explicitly demonstrate the comparison between our calculation and the currently available experimental data.
It is shown that the recent TOTEM data are well reproduced within our model, which strongly supports further applications of this framework.

The most important advantage of our model setup is its general versatility.
Since all the adjustable parameters contained in the model are fixed by the proton-proton data, other hadron-hadron processes at high energies can be analyzed without any additional parameter.
Hence, besides the nucleon-nucleon case, we also consider the pion-nucleon and pion-pion cases as examples in this work.
Similar to the nucleon case, the required density distribution of the pion in the AdS space can be obtained from the bottom-up AdS/QCD model of mesons~\cite{Erlich:2005qh,Abidin:2008hn}.
Although there is no available data for the pion involved processes in the considered kinematic region, we find that our prediction for the total cross section ratio $\sigma_{tot}^{\pi N} / \sigma_{tot}^{NN}$ agrees with that extracted from the well-known work with the soft Pomeron exchange done by Donnachie and Landshoff~\cite{Donnachie:1992ny}.

\section{Model setup}
\label{sec-2}
In this work, we investigate total cross sections of the high energy hadron-hadron scattering, and the formalism is in principle applicable for any hadron pair.
Considering the two-to-two process, $1+2 \to 3+4$, and following the preceding studies~\cite{Brower:2006ea,Brower:2007qh,Brower:2007xg,Brower:2010wf}, the scattering amplitude is written with the BPST Pomeron exchange kernel $\chi$ in the five-dimensional AdS space as
\begin{equation}
{\cal A} (s, t ) = 2 i s \int d^2 b e^{i \bm{k_\perp } \cdot \bm{b}} \int dz dz' P_{13} (z) P_{24} (z') \left[ 1 - e^{i \chi (s, \bm{b}, z, z' )} \right],  \label{eq:amp}
\end{equation}
where $s$ and $t$ denote the Mandelstam variables, $\bm{b}$ represents the two-dimensional impact parameter, and $z$ and $z'$ are fifth coordinates for the involved two hadrons.
$P_{13} (z)$ and $P_{24} (z')$ represent density distributions of the hadrons in the AdS space and are normalized, because the considered hadrons are normalizable modes.

Applying the optical theorem to Eq.~\eqref{eq:amp} and picking up the leading contribution from the eikonal representation, the expression for the total cross section can be obtained as
\begin{equation}
\sigma_{tot} (s) = 2 \int d^2 b \int dz dz' P_{13} (z) P_{24} (z') \mbox{Im} \chi (s, \bm{b}, z, z' ). \label{eq:tcs_1}
\end{equation}
In the conformal limit, it is known that the analytical form of the imaginary part of $\chi$ can be written down, and the impact parameter integration can be performed analytically.
As a result, Eq.~\eqref{eq:tcs_1} is rewritten as
\begin{align}
&\sigma_{tot} (s) = \frac{g_0^2 \rho^{3/2} }{8 \sqrt{\pi} } \int dz dz' P_{13} (z) P_{24} (z' ) (z z' ) \mbox{Im} [\chi_{c} (s, z, z' )],  \label{eq:tcs_2} \\
&\mbox{Im} [\chi_c (s, z, z' ) ] \equiv e^{(1 - \rho) \tau } e^ { - [ ({\log ^2 z / z'}) / {\rho \tau}]} / {\tau^{1/2}}, \label{eq:ck} \\
&\tau = \log (\rho z z' s / 2 ),
\end{align}
where $\rho$ and $g_0^2$ are adjustable parameters which control the energy dependence and the magnitude of the cross section, respectively.

Although the expressions shown above were rigorously derived, it is also known through the preceding applications of the BPST kernel to the DIS at small $x$ that the inclusion of the confinement effect is required to reproduce the structure function data measured at HERA.
Hence, we utilize the modified kernel, in which the added term mimics the confinement effect, for the numerical evaluations in this work, given by
\begin{align}
&\mbox{Im} [\chi_{mod} (s, z, z' )] \equiv \mbox{Im} [\chi_c (s, z, z' ) ] + \mathcal{F} (s, z, z' ) \mbox{Im} [\chi_c (s, z, z_0 z_0' / z' ) ],  \label{eq:mk} \\
&\mathcal{F} (s, z, z' ) = 1 - 2 \sqrt{\rho \pi \tau } e^{\eta^2 } \mbox{erfc} (\eta ),  \\
&\eta = \left( -\log \frac{z z' }{z_0 z_0' } + \rho \tau \right) / {\sqrt{\rho \tau }},
\end{align}
where the functional form of $\mbox{Im} [\chi_c ]$ is the same as that of Eq.~\eqref{eq:ck}.
Also, $z_0$ and $z'_0$ appearing here are the cutoffs in the fifth coordinates that characterize the QCD scale, but these are not adjustable because they are uniquely fixed by the hadron masses.

To numerically evaluate the total cross sections, it is also needed to specify the density distributions, in which the involved hadron properties reflect.
The density distributions are extracted from the hadron-Pomeron(graviton)-hadron three-point functions which can be computed by utilizing the bottom-up AdS/QCD models.
In this work, we evaluate the nucleon-nucleon, pion-nucleon, and pion-pion total cross sections, so we need the density distributions of the nucleon and the pion.
Since the authors of Refs.~\cite{Abidin:2008hn,Abidin:2009hr} calculated the gravitational form factors of those hadrons, using the classical actions of the AdS/QCD models of the nucleon~\cite{Hong:2006ta} and mesons~\cite{Erlich:2005qh}, we utilize their results in this study.
Those results were also applied to the analysis of the structure functions of the nucleon and the pion in DIS at small $x$ in the previous studies~\cite{Watanabe:2012uc:Watanabe:2013spa}.

\section{Numerical results}
\label{sec-3}
There are in fact some parameters included in the bottom-up AdS/QCD models which are used to calculate the required density distributions of the nucleon and the pion.
However, they can be fixed with the hadronic observables such as the masses and the decay constants.
Therefore, the adjustable parameters to be determined with the experimental data here are the two, $\rho$ and $g_0^2$, which are contained in the BPST Pomeron exchange kernel as seen in the previous section.
Besides the proton-proton collision data recently measured by the TOTEM collaboration at LHC~\cite{Antchev:2017dia,Antchev:2013gaa:Antchev:2013iaa:Antchev:2013paa:Antchev:2015zza:Antchev:2016vpy:Nemes:2017gut}, other pp~\cite{Baltrusaitis:1984ka:Honda:1992kv:Collaboration:2012wt}
and
$\bar{\rm{p}}$p~\cite{Battiston:1982su:Hodges:1983oba:Bozzo:1984rk:Alner:1986iy:Amos:1991bp:Abe:1993xy:Augier:1994jn:Avila:2002bp} data, which were summarized in 2010 by the Particle Data Group~\cite{Nakamura:2010zzi}, are considered to determine those parameters in this study.

\begin{figure}[bt!]
\centering
\includegraphics[width=0.85\textwidth]{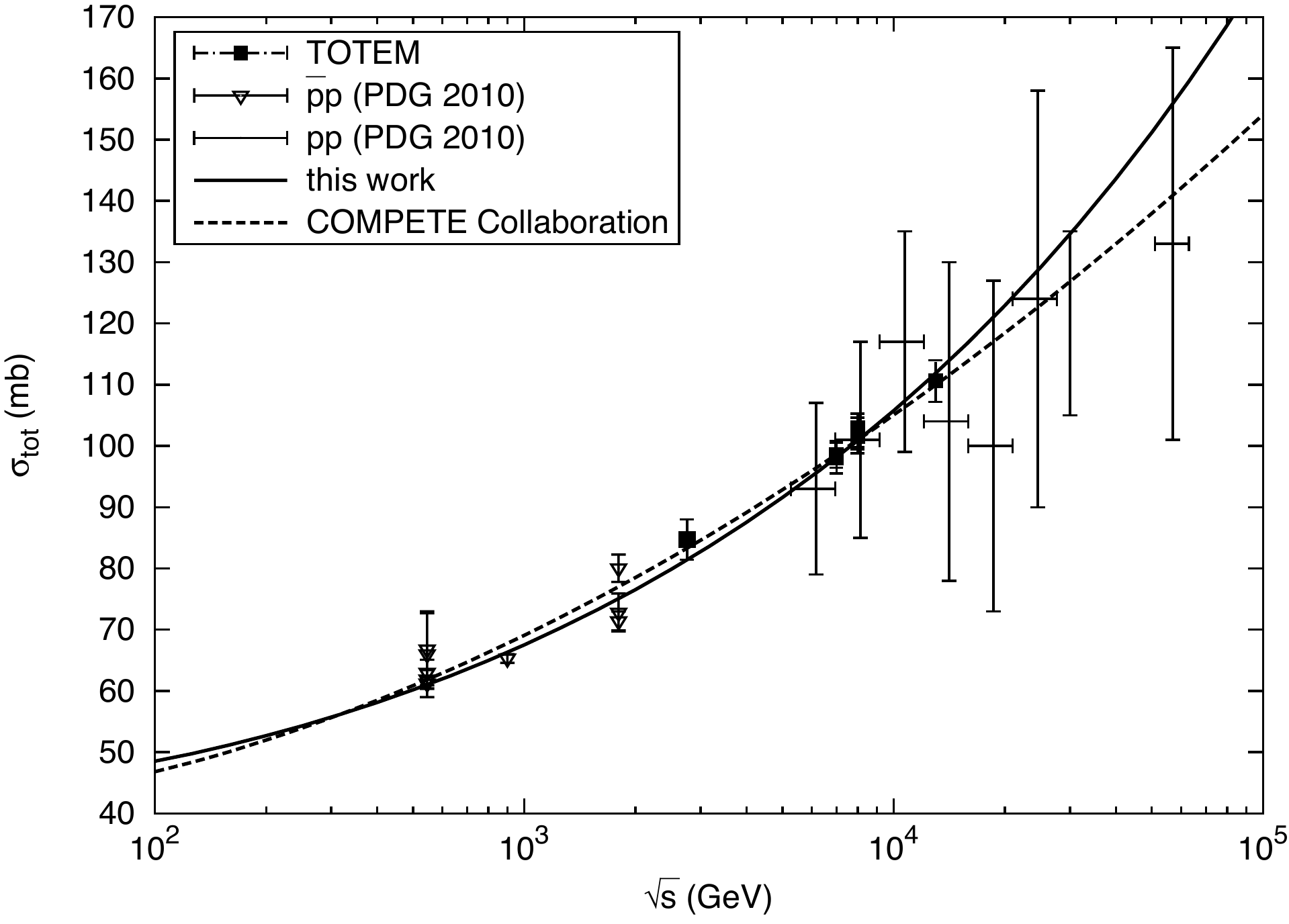}
\caption{
The nucleon-nucleon total cross sections as a function of $\sqrt{ s }$. 
The solid and dashed lines depict our calculation and the empirical fit by the COMPETE collaboration~\cite{Cudell:2002xe}, respectively.
The recent TOTEM data~\cite{Antchev:2017dia,Antchev:2013gaa:Antchev:2013iaa:Antchev:2013paa:Antchev:2015zza:Antchev:2016vpy:Nemes:2017gut}
and other
pp~\cite{Baltrusaitis:1984ka:Honda:1992kv:Collaboration:2012wt}
and
$\bar{\rm{p}}$p~\cite{Battiston:1982su:Hodges:1983oba:Bozzo:1984rk:Alner:1986iy:Amos:1991bp:Abe:1993xy:Augier:1994jn:Avila:2002bp} data are plotted with error bars.
}
\label{fig:TCS_nucleon}
\end{figure}

We display the resulting nucleon-nucleon total cross section in Fig.~\ref{fig:TCS_nucleon},
in which, besides the experimental data, we also plot the empirical fit obtained by the COMPETE collaboration~\cite{Cudell:2002xe} for comparison.
It is seen in Fig.~\ref{fig:TCS_nucleon} that our calculation is in agreement with the data in the whole considered region, and also is consistent with the empirical fit at $\sqrt{s} < 10$~TeV.
In the higher energy region, our result becomes larger than the empirical one, and the size of the deviation between the two curves increases with $\sqrt{s}$.

\begin{figure}[bt!]
\centering
\includegraphics[width=0.55\textwidth]{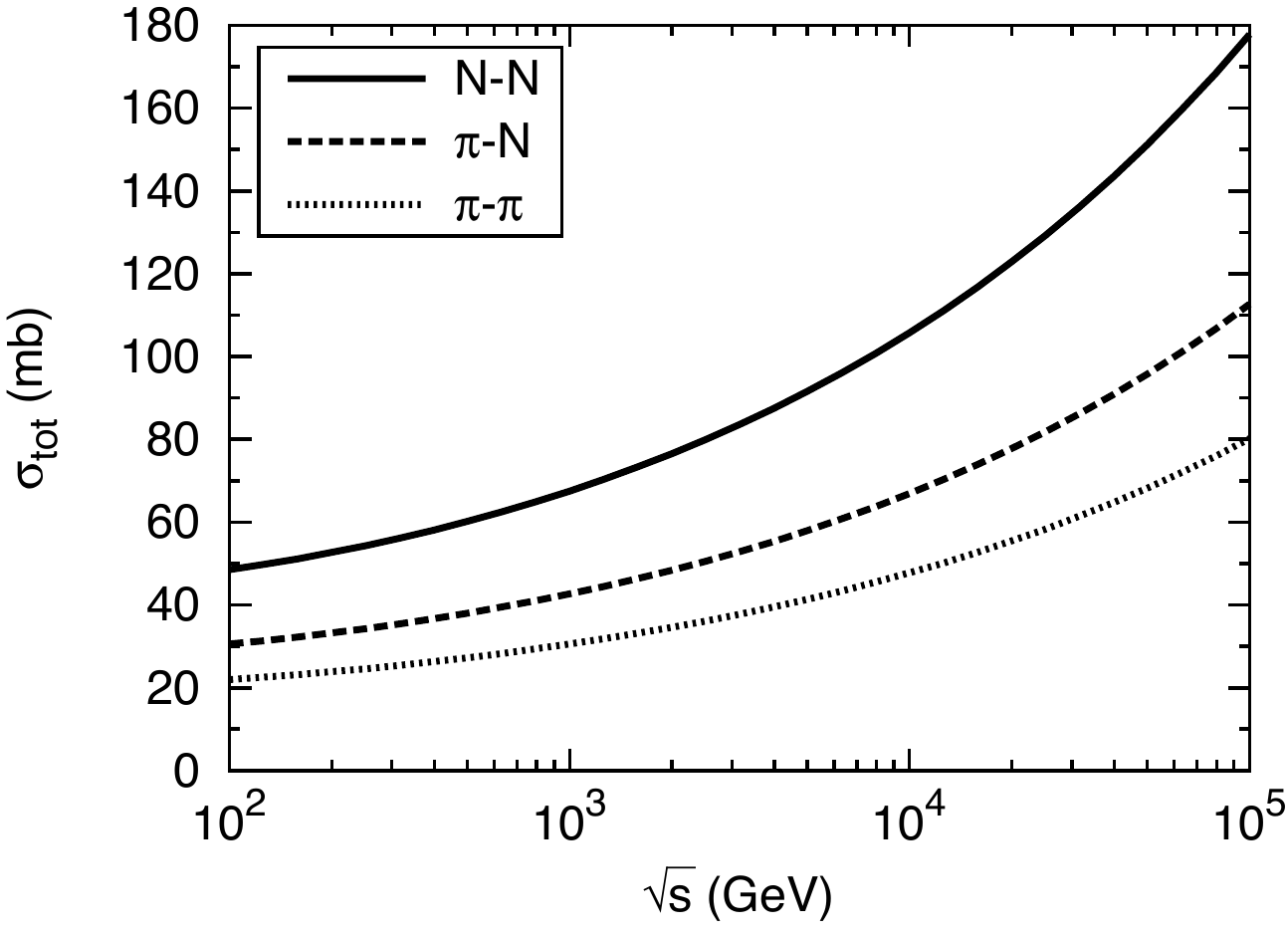}
\caption{
The total cross sections obtained in this work as a function of $\sqrt{s}$.
The solid, dashed, and dotted lines depict the nucleon-nucleon, pion-nucleon, and pion-pion results, respectively.
}
\label{fig:TCS_pion}
\end{figure}

Next, we show in Fig.~\ref{fig:TCS_pion}
our results for the pion-nucleon and pion-pion total cross sections.
These cross sections can be calculated without any adjustable parameter, because all the parameters are fixed with the pp($\bar{\rm{p}}$p) data.
One can see from the figure that the pion-nucleon result is substantially smaller than the nucleon-nucleon one, and the pion-pion result is smallest among the three curves.
Since in our model the energy dependence is only from the BPST kernel, the $s$ dependence is common to the three results, which implies the universality of the BPST Pomeron.
Therefore, we can predict the total cross section ratios as follows:
\begin{equation}
\frac{\sigma_{tot}^{\pi N }}{\sigma_{tot}^{N N }} = 0.63, \ \
\frac{\sigma_{tot}^{\pi \pi }}{\sigma_{tot}^{N N }} = 0.45.
\end{equation}
Although there is no experimental data for the pion involved processes, an empirical value can be extracted from the work with the soft Pomeron exchange done by Donnachie and Landshoff~\cite{Donnachie:1992ny}.
The extracted ratio is $\sigma_{tot}^{\pi N } / \sigma_{tot}^{N N } = 0.63$ which agrees with our prediction.

\section{Summary}
\label{sec-4}
In this work, we have investigated the hadron-hadron total cross sections at high energies in the framework of holographic QCD whose formalism is constructed in the five-dimensional AdS space.
Applying the BPST Pomeron exchange kernel to describe the involved strong interaction and utilizing the bottom-up AdS/QCD models to obtain the density distributions of the hadrons, we have calculated the total cross sections.

Our nucleon-nucleon result is in agreement with the experimental data, including the recently measured TOTEM ones.
Our calculation is also consistent with the empirical fit obtained by the COMPETE collaboration at $\sqrt{s} < 10$~TeV, although a substantial deviation is observed at higher $s$.
Since the currently available data at $\sqrt{s} > 10$~TeV, except for the recent TOTEM's 13~TeV one, were extracted from the cosmic-ray experiments, they have huge uncertainties.
Hence, more precise data in such a very high $s$ regime are needed to pin down the total cross section.

In our model setup, we can consider various hadron pairs as the participants in the hadron-hadron scattering, giving appropriate density distributions for them.
Besides the nucleon-nucleon case, we have also analyzed the pion-nucleon and pion-pion cases, and predicted the total cross section ratios.
Our predictions could be tested at future experimental facilities with a high intensity pion beam.

Our framework is in principle applicable for any high energy scattering process if the strong interaction involved in the process can be approximated by the Pomeron exchange.
Further investigations are certainly needed.

\section*{Acknowledgments}
\label{acknowledgments}
This work was supported by the NSFC under Grant Nos.
11725523,
11735007,
and
11261130311
(CRC 110 by DFG and NSFC).

\end{document}